\documentclass[aip,cha,twocolumn,showpacs,preprintnumbers,amsmath,amssymb,superscriptaddress,10pt]{revtex4-1}

\usepackage{amsmath,amssymb}
\usepackage{bm}
\usepackage[latin1]{inputenc}
\usepackage{dcolumn}
\usepackage{graphicx}
\usepackage{SIunits}
\usepackage{multirow}
\usepackage{ae}
\usepackage{subfigure}
\usepackage[top=1.5cm, bottom=1.5cm, left=1.6cm, right=1.6cm]{geometry}

\begin{document}

\title{\Large{Reply to ``Avoiding the Detector Blinding Attack on Quantum Cryptography''}}

\author{Lars Lydersen}
\affiliation{Department of Electronics and Telecommunications, Norwegian University of Science and Technology, NO-7491 Trondheim, Norway}
\affiliation{University Graduate Center, NO-2027 Kjeller, Norway}

\author{Carlos Wiechers}
\affiliation{Max Planck Institute for the Science of Light, G\"{u}nther-Scharowsky-Strasse 1/Bau 24, 91058 Erlangen, Germany}
\affiliation{Institut f\"{u}r Optik, Information und Photonik, University of Erlangen-Nuremberg, Staudtstra\ss e 7/B2, 91058 Erlangen, Germany}
\affiliation{Departamento de F\'{i}sica, Universidad de Guanajuato, Lomas del Bosque 103, Fraccionamiento Lomas del Campestre, 37150, Le\'{o}n, Guanajuato, M\'{e}xico}

\author{Christoffer Wittmann}
\author{Dominique Elser}
\affiliation{Max Planck Institute for the Science of Light, G\"{u}nther-Scharowsky-Strasse 1/Bau 24, 91058 Erlangen, Germany}
\affiliation{Institut f\"{u}r Optik, Information und Photonik, University of Erlangen-Nuremberg, Staudtstra\ss e 7/B2, 91058 Erlangen, Germany}

\author{Johannes Skaar}
\affiliation{Department of Electronics and Telecommunications, Norwegian University of Science and Technology, NO-7491 Trondheim, Norway}
\affiliation{University Graduate Center, NO-2027 Kjeller, Norway}

\author{Vadim Makarov}
\affiliation{Department of Electronics and Telecommunications, Norwegian University of Science and Technology, NO-7491 Trondheim, Norway}


\maketitle

\thispagestyle{empty}

{\bf This is a reply to the comment\cite{yuan2010} by Yuan \emph{et al}$.$ on our publication\cite{lydersen2010a}.}

We are glad that our results have led to awareness and discussions of imperfections in the detectors, among the leading research groups on avalanche-photodiode-based (APD-based) quantum key distribution (QKD) systems. Yuan \emph{et al$.$} propose a method \cite{yuan2010} to avoid blinding of gated APD-based detectors, such as the ones used in the two commercial QKD systems addressed in our recent publication \cite{lydersen2010a}. Our experimental data from Clavis2 indicate that the countermeasure suggested by Yuan \emph{et al.} will make it more difficult to blind gated detectors.

However, for gated detectors avoiding blinding is insufficient to avoid our attack. Gated detectors operate in linear mode between the gates. Therefore the trigger pulse can be applied right after the gate (discarding these clicks based on arrival times seems to be impractical due to detector jitter). We remarked that this causes afterpulses \cite{lydersen2010a}, but in fact the after-gate attack can fully compromise the security for a wide range of system parameters \cite{wiechers2010}. Even outside this range, one must quantify in a proof of security how well Eve may perform. Removing the bias resistor and lowering the comparator threshold does not avoid exploiting the linear mode between gates. In fact, lowering the comparator threshold reduces the required trigger pulse power, and thus likely improves the after-gate attack by reducing afterpulsing.

Furthermore, it seems that the detectors can still be blinded, even with the changes proposed by Yuan \emph{et al.} Simply removing the bias resistor has turned out to be insufficient. In our recent paper \cite{lydersen2010b}, we removed the bias resistor from the detectors in Clavis2. Still, we were able to blind the detectors in several ways. Yuan \emph{et al$.$} did not observe thermal blinding from  continuous-wave (c.w) illumination. This may be due to the lower comparator threshold and/or insufficient heating (they illuminate one instead of two APDs, while operating at a higher temperature, effectively increasing cooler capacity).

Even if the bias resistor is removed, \emph{and} the discrimination level is set just above the capacitive charging signal the detectors seem to be vulnerable to \emph{sinkhole} blinding \cite{lydersen2010b}. In sinkhole blinding, the APD is illuminated between the gates. With a suitable duty cycle of the blinding illumination, it should be straightforward to blind the detector while keeping the comparator input well below the amplitude of the capacitive signal.

Monitoring photocurrent of the APDs is like using a power meter at Bob's entrance, which we have already discussed in our original paper \cite{lydersen2010a}. Furthermore, it will not reveal the after-gate attack.

It seems that the countermeasure proposed by Yuan \emph{et al$.$} \cite{yuan2010} does not prevent our general attack of tailored bright illumination \cite{lydersen2010a}. Note that so far, we have been able to blind and control every APD-based detector which we have looked thoroughly at (although with different techniques), including three different passively quenched detectors \cite{makarov2009}, one actively quenched detector \cite{makarov2008a}, and two different gated detectors \cite{lydersen2010a,lydersen2010b,wiechers2010}.

In our opinion, the discussion shows how important it is to close this loophole in a thorough, preferably provable way. We doubt that this could be achieved efficiently in small increments of intuitive patches, causing rapid iterations and forcing manufacturers to update their QKD systems frequently. We are confident that APD-based single photon detectors can be, and will be made secure by a proper implementation combined with a sufficiently general security proof.

As a final remark, we want to emphasize that in our experiments \cite{lydersen2010a,lydersen2010b,wiechers2010}, the QKD systems were treated as black boxes just as they would be for Eve. We reverse-engineered the detector circuitries (realistically, Eve can buy a copy of Bob and do the same), and non-intrusively recorded the detector response during our experiments. Clavis2 shipped with factory settings ready for QKD, including the discrimination level, which we used for our experiments. As pointed out in our Supplementary Information\cite{lydersen2010a}, QPN 5505 did not ship with factory settings, but we followed the manual and used the settings which gave us the best QKD performance.



\end{document}